\documentstyle[]{mn}

\title[]{{\sl XMM-Newton} observations of the eclipsing polar V2301 Oph}

\author[Ramsay \& Cropper]{
Gavin Ramsay\thanks{Present Address: Armagh Observatory, College Hill,
Armagh, Northern Ireland, gar@arm.ac.uk} and Mark Cropper\\
Mullard Space Science Laboratory, University College London,
Holmbury St. Mary, Dorking, Surrey, RH5 6NT, UK\\
}
\date{Received: }

\begin{document}
\outer\def\gtae {$\buildrel {\lower3pt\hbox{$>$}} \over 
{\lower2pt\hbox{$\sim$}} $}
\outer\def\ltae {$\buildrel {\lower3pt\hbox{$<$}} \over 
{\lower2pt\hbox{$\sim$}} $}
\newcommand{\gscm}  {g s$^{-1}$ cm$^{-2}$} 
\newcommand{\ergscm} {ergs s$^{-1}$ cm$^{-2}$}
\newcommand{\ergss} {ergs s$^{-1}$}
\newcommand{\ergsd} {ergs s$^{-1}$ $d^{2}_{100}$}
\newcommand{\pcmsq} {cm$^{-2}$}
\newcommand{\ros} {\sl ROSAT}
\newcommand{\exo} {\sl EXOSAT}
\newcommand{\xmm} {\sl XMM-Newton}
\def\rchi{{${\chi}_{\nu}^{2}$}}
\def\uchi{{${\chi}^{2}$}}
\newcommand{\Msun} {$M_{\odot}$}
\newcommand{\Mwd} {$M_{wd}$}
\def\Mdot{\hbox{$\dot M$}}
\def\mdot{\hbox{$\dot m$}}

\maketitle

\begin{abstract}

We present {\sl XMM-Newton} observations of the eclipsing polar V2301
Oph which cover nearly 2.5 binary orbital cycles and 2 eclipses. This
polar is believed to have the lowest magnetic field strength (7 MG) of
any known polar. We find evidence for structure in the X-ray eclipse
profile which shows a `standstill' feature lasting 26$\pm$4 sec. This
allows us to place an upper limit on the mass of the white dwarf of
$\sim$1.2 \Msun. We find no evidence for QPOs in the frequency range
0.02-10 Hz.  This coupled with the absence of QPOs in {\sl RXTE} data
suggest that, if present, any oscillations in the shock front have a
minimal effect on the resultant X-ray flux. We find no evidence for a
distinct soft X-ray component in its spectrum - it therefore joins
another 7 systems which do not show this component. We suggest that
those systems which are asynchronous, have low mass transfer rates, or
have accretion occurring over a relatively large fraction of the white
dwarf are more likely to show this effect. We find that the specific
mass transfer rate has to be close to 0.1 g cm$^{-2}$ s$^{-1}$ to
predict masses which are consistent with that derived from our eclipse
analysis. This maybe due to the fact that the low magnetic field
strength allows accretion to take place along a wide range of azimuth.

\end{abstract}


\section{Introduction}

Polars are binary systems in which the accreting white dwarf has a
sufficiently high magnetic field strength to prevent the formation of
an accretion disc. The magnetic field strength of the white dwarf lies
in the range $\sim$7-200MG. The system with the lowest magnetic field
strength is believed to be the eclipsing system V2301 Oph (Ferrario et
al 1995).

V2301 Oph has been observed in X-rays using {\ros}, where it was
bright in X-rays and showed a relatively hard spectrum (Ramsay 1997,
Hessman et al 1997); {\sl RXTE} (Steinman-Cameron \& Imamura 1999) and
{\sl ASCA} (Terada et al 2004). In the optical band a bright accretion
stream is evident and there is a significant cycle-to-cycle variation
in its eclipse profile. Reynolds et al (2005) suggest that this
variation is due to a change in the amount of material in the
threading region, the point at which the accretion stream feels the
force of the magnetic field of the white dwarf.

Because V2301 Oph is bright in X-rays and also an eclipsing binary
(113 min, Silber et al 1994) it is an excellent source to study the
accretion process in polars. In this paper we present an analysis of
data taken using {\xmm} made in Sept 2004. V2301 Oph was the last
source to be observed as part of the {\xmm}-MSSL polar survey which
provided snap-shots of nearly half of the known systems (Ramsay \&
Cropper 2004, Ramsay et al 2004b).

\section{Observations and Data Reduction} 

V2301 Oph has been observed using {\it XMM-Newton} on 5 separate
occasions.  The first 4 planned observations were compromised by high
background radiation, which resulted in the observations not being
carried out, or exposures which were short: we do not consider these
observations further. However, the observation which took place on
Sept 2004 was not affected by high background radiation. The details
of the exposure time in the different instruments are shown in Table
\ref{log}. The EPIC MOS (Turner et al 2001) start time is earlier (and
the observations slightly longer) than that of the EPIC-pn
(Str\"{u}der et al 2001). Observations with the Optical Monitor (Mason
et al 2001) were carried out in fast mode and in two filter bands: the
$V$-band, and UVW1 (2400$\--$3400\,\AA) band. In the optical the mean
brightness was $V$=17.5, indicating it was in a relatively high
accretion state -- its observed range is $V\sim16-21$ (Warner 1999).

\begin{figure*}
\begin{center}
\setlength{\unitlength}{1cm}
\begin{picture}(12,10.3)
\put(-1.6,0.4){\includegraphics{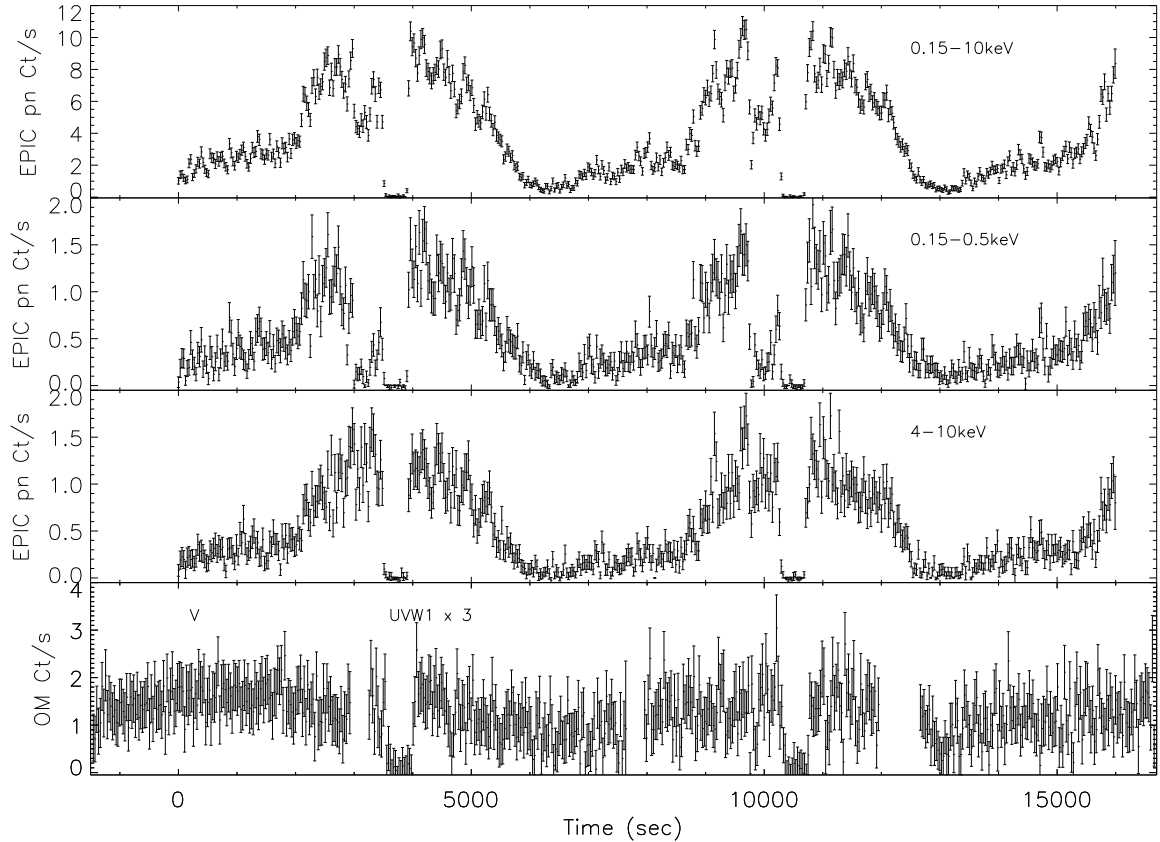}}
\end{picture}
\end{center}
\caption{The unfolded EPIC pn and OM light curves for V2301 Oph. The
data has been binned into 30 sec bins. From the top we show the
light curves in the: 0.15-10keV, 0.15-0.5keV, 4-10keV, V band and UVW1 band.}
\label{unfolded}
\end{figure*}

The data were reduced using {\tt SAS} v7.0. Only X-ray events which
were graded as {\tt PATTERN}=0-4 and {\tt FLAG}=0 were used. Events
were extracted from a circular aperture centered on the source, with
background events being extracted from a source free area. The
background data were scaled to give the same area as the source
extraction area and subtracted from the source area.  The events in
the EPIC pn detector were slightly piled-up during the time intervals
which had the greatest count rate. Since the MOS detectors were
operating in small window mode, the MOS data were not piled-up. The
RGS data were reduced using {\tt rgsproc} and the first order spectra
from the two RGS detectors were extracted. The OM data were reduced
using {\tt omfchain}.

\begin{table}
\begin{center}
\begin{tabular}{llcr}
\hline
Instrument & Mode & Filter & Duration\\
\hline
EPIC MOS & Small window & Thin & 18657\,s\\
EPIC PN    & Large window   & Thin & 16657\,s\\
RGS        & Spectro + Q          &  & 18885\,s\\
OM         & Fast Mode & V          & 4399\,s\\
OM         & Fast Mode & UVW1   & 13000\,s\\
\hline
\end{tabular}
\caption{The summary for {\xmm} observations of V2301 Oph taken in
2004 Sept 6.}
\label{log}
\end{center}
\end{table}

\section{The Light Curves}

\subsection{General Features}

The peak count rate in the EPIC pn detector is $\sim$10 ct/s in the
0.15-10keV band, making V2301 Oph one of the brightest of polars in
this band. Approximately 2.5 orbital cycles and two eclipses were
observed (Figure \ref{unfolded}). X-ray flux is seen throughout the
orbital cycle, implying that at least one accretion region is seen at
all times.

\begin{figure}
\begin{center}
\setlength{\unitlength}{1cm}
\begin{picture}(6,12.7)
\put(-1.6,-0.7){\includegraphics{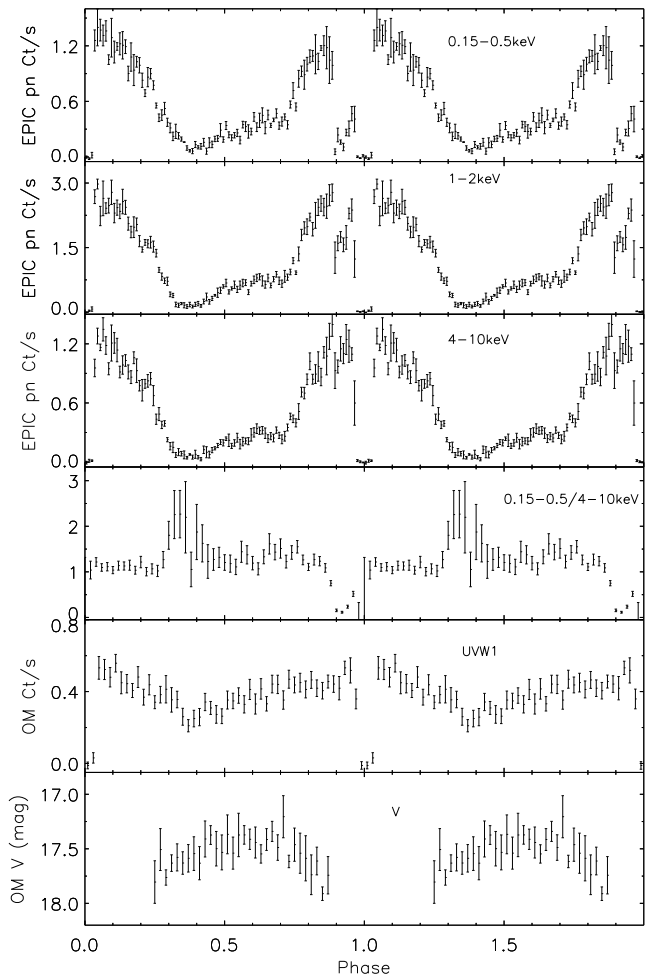}}
\end{picture}
\end{center}
\caption{The EPIC pn and OM light curves for V2301 Oph folded on the
orbital ephemeris of Barwig et al (1994). The top three panels have
been binned into 100 bins, while the rest have been binned into 50
bins. From the top we show the folded light curves in the:
0.15-0.5keV, 1-2keV, 4-10keV, 0.15-0.5keV/4-10keV and the UV energy
bands.}
\label{light_fold}
\end{figure}

We show the light curves folded on the ephemeris of Barwig et al
(1994) (after placing it onto TT) in Figure \ref{light_fold}. As
already noted by Hessman et al (1997), a dip is seen in the X-ray
light curve at $\phi$=0.9 which is thought to result from the
accretion stream obscuring the bright accretion region -- this is
manifest in the softness ratio which becomes significantly harder at
this phase. A short dip is also seen in the unfolded UV light curve at
this phase (Figure \ref{unfolded}). After the eclipse, at
$\phi\sim$0.3--0.4, the softness ratio becomes significantly
softer. This could be due to one of the accretion regions having
rotated out of view, with the region which is still visible having a
much softer X-ray spectrum. The UV light curve shows a
quasi-sinusoidal light curve and is likely to be due to the result of
one or more hotter areas on the white dwarf surface rotating in and
out of view.

\subsection{The ephemeris}
\label{ephemeris}

To calculate the time of mid-eclipse we determined the time of the
start of the steep decline in the X-ray flux by eye and the end of
the steep increase in X-ray flux. We defined the mid-point of these
times as the time of mid-eclipse -- we report these times in Table
\ref{eclipse_times}.

\begin{figure}
\begin{center}
\setlength{\unitlength}{1cm}
\begin{picture}(6,5.5)
\put(-1.2,-0.2){\includegraphics{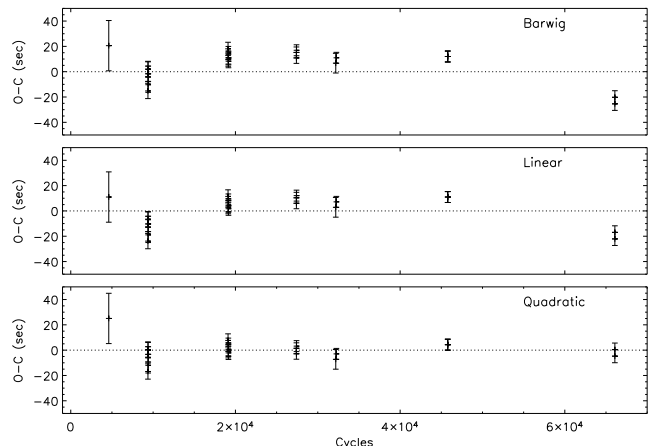}}
\end{picture}
\end{center}
\caption{The O-C diagram for the eclipse timings of V2301 Oph. The last 
2 points are those determined from {\xmm} data. The top panel shows
the residuals using the ephemeris of 
Barwig et al (1994) (after placing it on to the TT time system); the middle
panel shows the residuals using a new linear ephemeris; the bottom
panel shows the residuals to the quadratic ephemeris reported in the text.}
\label{o_c}
\end{figure}

Compared to the ephemeris of Barwig et al (1994), the mid-point of the
eclipses are earlier by 20--25 sec. To determine if there has been a
systematic change in the arrival time of the eclipse, we obtained all
the eclipse times from the literature. Some of these timings were not
accurate enough to include and there appeared to be some typographical
errors in 2 timings (the 5th and 6th timings reported in
Steiman-Cameron \& Imamura 1999). We corrected each time so that they
were on the TT time system (ie including the shift from UTC to TAI and the
appropriate number of leap seconds). We used the ephemeris of Barwig
et al (1994) (after placing it onto TT) to phase the eclipse times and
the residuals are shown in the upper panel of Figure \ref{o_c}.

To determine if we could derive an improved ephemeris, we fitted all
the eclipse times with a linear ephemeris and also an ephemeris with a
quadratic term. Using these ephemerides we calculated the goodness of
fit for each using the chi-squared estimator. In many of the previous
eclipse time studies we had to estimate the uncertainty in the
mid-eclipse time since they were not explicitly stated. The goodness
of fit using the Barwig et al (1994) ephemeris was \rchi=7.1 (21 dof);
using our linear ephemeris the fit was \rchi=5.23 (21 dof); and using
our quadratic ephemeris the fit was \rchi=1.17 (20 dof). Our quadratic
ephemeris:

T$_{eclipse}$ (TT) = \newline 2448071.020690(61) + 0.0784500274(44) - 
$3.18(62)\times 10^{-13}  E^{2}$

gives a better fit than our linear ephemeris with a confidence level
of greater than 99.9\%.  The quadratic term could be due to an
intrinsic change in the orbital period, an asynchronicity between the
orbital period and the spin period of the white dwarf or a secular
shift in the location of the accretion region(s) on the white
dwarf. We note that the magnitude of the quadratic term is similar to
that observed in HU Aqr, Schwope et al (2001), who suggested that the
quadratic term seen HU Aqr was due to a shift in the location of the
X-ray emission regions on the surface of the white dwarf.

\begin{table}
\begin{center}
\begin{tabular}{lrr}
\hline
Mid-Eclipse   & Error  & Reference \\
(TT) & (Days) & \\
\hline
2448432.98940 &  0.00023 &(1) \\
2448806.48947 &  0.00007 &(2) \\
2448806.56813 &  0.00007 &(2) \\
2448806.64653 &  0.00007 &(2) \\
2448807.50952 &  0.00007 &(2) \\
2448808.60768 &  0.00007 &(2) \\
2448809.39228 &  0.00007 &(2) \\
2449569.41619 &  0.00006 &(3) \\
2449571.45579 &  0.00006 &(3) \\
2449572.39725 &  0.00006 &(3) \\
2449572.55411 &  0.00006 &(3) \\
2449573.41705 &  0.00006 &(3) \\
2449573.49547 &  0.00006 &(3) \\
2450220.78652 &  0.00005 &(4) \\
2450220.86502 &  0.00005 &(4) \\
2450221.80644 &  0.00005 &(4) \\
2450595.77751 &  0.00009 &(4) \\
2450601.73976 &  0.00005 &(4) \\
2451661.67783 &  0.00005 &(5) \\
2451662.61923 &  0.00005 &(5) \\
2451663.56063 &  0.00005 &(5) \\
2453254.68332 &  0.00006 &(6) \\
2453254.76171 &  0.00006 &(6) \\
\hline
\end{tabular}
\caption{The observed times for the mid-eclipse of V2301 Oph.  To
ensure the times are on a common reference frame, we have added to the
times reported in the literature (where necessary) 32.184 sec to
convert UTC times on to TAI and also the appropriate number of leap
seconds. For those references which did not explicitly give errors on
the mid-eclipse times, we have estimated the error for that point
using the available information. References: (1) Silber et al (1994);
(2) Barwig et al (1994); (3) \v{S}{i}mi\'{c} et al (1998); (4)
Steiman-Cameron \& Imamura (1999); Reynolds et al (2005); (6) this
paper.}
\label{eclipse_times}
\end{center}
\end{table}

\subsection{The eclipse}
\label{eclipse}

\begin{figure}
\begin{center}
\setlength{\unitlength}{1cm}
\begin{picture}(6,10)
\put(-0.8,-1){\includegraphics{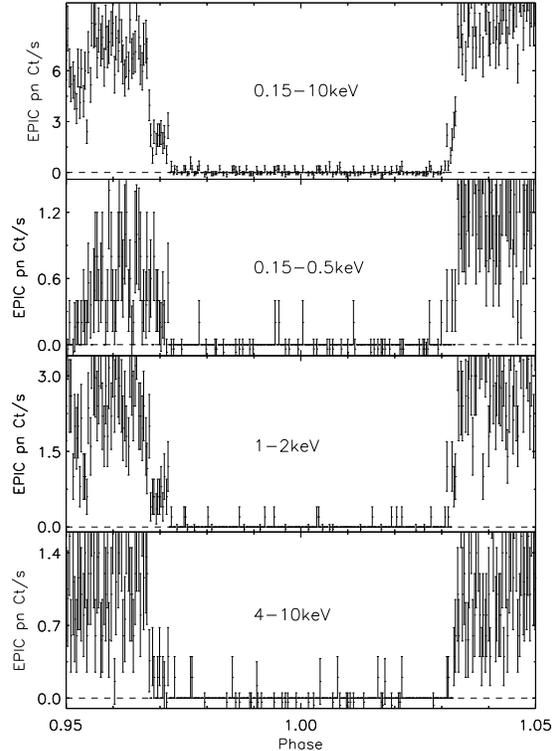}}
\end{picture}
\end{center}
\caption{The X-ray light curves in 4 energy bands extracted from EPIC
pn data folded on the quadratic ephemeris shown in \S \ref{ephemeris}.
The time bins are 5 sec and the plot shows data from 2 eclipses. The
`standstill' which is visible in each light curve apart from the
0.15--0.5keV curve, has a duration of 26$\pm$4sec sec. The dashed line 
indicates 0.0 cts/s.}
\label{eclipse_energy}
\end{figure}

In Figure \ref{eclipse_energy} we show the X-ray light curve in 4
energy bands where we have folded the light curves on our quadratic
ephemeris. The descent into eclipse is very rapid ($<$5 sec) which is
followed by a `standstill' lasting 26$\pm$4 sec (determined by eye)
in the light curves made using photons with energies greater than
0.5keV.  This is likely due to one accretion region, and then a
second, being eclipsed by the secondary star. The UV light curve does
not show this feature. After the eclipse of the second accretion
region there is a rapid ($<$5 sec) descent into total eclipse. During
the eclipse egress (which lasts 23$\pm$2sec), there is no evidence for
a standstill.

As far as we can determine, this is the first time that such a
standstill has been observed in the X-ray light curve of a
polar. However, a standstill feature in both the eclipse ingress and
egress has been seen in the optical band in UZ For (Perryman et al
2001) and SDSS J015543.40+002807.2 (O'Donoghue et al 2006).  If X-rays
from the accretion regions are emitted close to the surface of the
white dwarf (Cropper, Wu \& Ramsay 2000), the time duration of this
standstill puts an upper limit on the mass of the white dwarf,
$M_{1}$.

To determine this upper limit, we assume a Roche Lobe geometry and the
Nauenberg (1972) mass-radius relationship for white dwarfs. We trace
the Roche potential out of the binary system along the line of sight
from any point in the vicinity of the white dwarf. As a starting point
we take the results of an analysis of optical observations of the
eclipse profile by Reynolds et al (2005) who favoured a mass ratio of
$q=0.15$ and an orbital inclination of $i=84^{\circ}$. The first
accretion region is eclipsed at $\phi$=0.9680 (Figure
\ref{eclipse_energy}) - this fixes the location of this region along
an arc defined by the limb of the secondary.  We then place the second
accretion region on the limb of the white dwarf so that it is the last
part of the white dwarf to be eclipsed (at $\phi$=0.9718). We find
that this sets an {\sl upper} limit of $M_{1}$=1.2 \Msun. (For a fixed
$q$ this upper limit is constrained to within $\sim$0.02 \Msun).

Silber et al (1994) found that the optical spectrum of V2301 Oph taken
at mid-eclipse was most similar to an M6~V spectral type. Based on the
theoretical models of Kolb \& Baraffe (2000), such a spectral type
implies a secondary star mass, $M_{2}=$0.10\Msun. The same models also
predict the spectral type of the secondary star as a function of
orbital period. For an orbital period appropriate to V2301 Oph (113
min) the models predict a secondary of spectral type of $\sim$M4.5
(Baraffe \& Kolb 2000), which implies $M_{2}$=0.15\Msun (Kolb \& Baraffe
2000). For $M_{2}$=0.10 and 0.15 \Msun and $q$=0.15 (Reynolds et al
2005), this implies $M_{1}$=0.67 and 1.0 \Msun respectively.

For $M_{1}$=1.0 \Msun, $M_{2}$=0.15 \Msun and $i=84^{\circ}$, we find
that we can reproduce the observed duration of the standstill
feature. Further, both accretion regions reappear simultaneously after
eclipse (and hence no standstill is observed) if the accretion regions
are located at (90$^{\circ}$,10$^{\circ}$) and
(60$^{\circ}$,45$^{\circ}$) where the co-ordinates are $\beta, \zeta$
(as defined by Cropper 1989). The shape of the accretion regions are
expected to be more complex than simple circular regions, but we have
found a solution which is self consistent and in agreement with the
results of Reynolds et al (2005).

\subsection{Searching for QPOs}

It is predicted that Quasi Periodic Oscillations (QPOs) may be seen on
a timescale of a few seconds in the light curves of AM Her systems if
the accretion shock front is not stable (eg Saxton \& Wu 1999, Mignone
2006 and references therein). For systems with high magnetic field
strengths, strong cyclotron cooling would tend to suppress these
QPO's. Since V2301 Oph is believed to have the lowest magnetic field
strength of any polar (Ferrario et al 1995) it is a good candidate for
detecting QPOs.

We searched for QPOs in the X-ray light curve. We extracted events in
the energy range 2--10keV (since these X-rays are expected to be
emitted closer to the shock than soft X-rays) from various time
intervals corresponding to both accretion regions. We searched for
QPO's using the event data and also in the light curves binned up into
0.1 sec bins in the frequency range 0.02-10 Hz using a Discrete
Fourier Transform (DFT). We do not detect any evidence for peaks in
the DFT above the noise level in this frequency range. We also
searched for QPOs at softer X-rays (0.15--1.0keV) and again found no
evidence for QPOs at these energies either.

To place constraints on the upper limit, we added a coherent signal of
varying amplitude to the binned light curves. We found that we could
not detect signals which had full amplitudes corresponding to 19--24\%
of the signal in the 2--10keV energy band.  Steiman-Cameron \& Imamura
(1999) also found no evidence for QPO's in either optical or X-ray
data {\sl RXTE} (2--60 keV). In X-rays they gave upper limits of
4.1--7.6\% which is significantly lower than our limits. We believe
that Steiman-Cameron \& Imamura (1999) used data from all orbital
phases in their analysis and they had a factor of $\sim$30 more counts
compared to our 2--10keV band light curves.

The predicted modulation amplitude is sensitive to many parameters
such as magnetic field strength, specific mass transfer rate, physical
size and shape of the accretion region, (not to mention the energy
band which is being used to measure the flux), so it difficult to
compare the observed limits to predictions.  However, these results
suggest that even if there is an oscillation in the shock front it
does not affect the observable X-ray flux - at energies which either
{\xmm} or {\sl RXTE} are sensitive.

\section{Spectra}

\subsection{RGS spectrum}
\label{rgs}

V2301 Oph is bright enough in X-rays to us allow us to extract a phase
averaged spectrum which has a moderate signal-to-noise ratio from the
RGS detectors.  We excluded data from a short time interval (550 sec)
where the background was significantly higher compared to the rest of
the observation.

\begin{figure*}
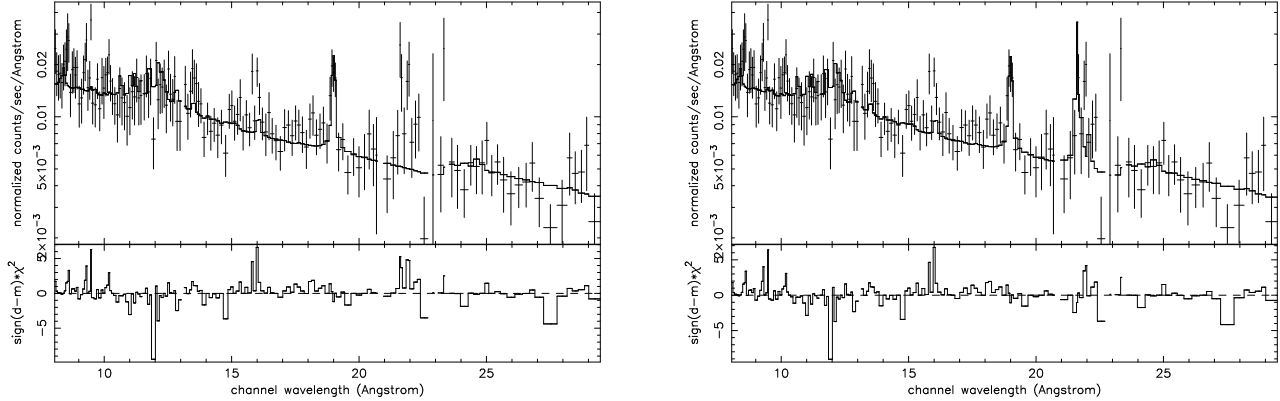

\begin{center}
\setlength{\unitlength}{1cm}
\begin{picture}(6,5)
\put(-5.5,-0.5){\includegraphics{wasacg_mat.ps}}
\put(3.5,-0.5){\includegraphics{wasacggau_mat.ps}}
\end{picture}
\end{center}
\caption{The X-ray spectrum of V2301 Oph taken using the RGS (the data
taken using the RGS1 and RGS2 detectors have been combined). In the
left hand panel we show the fit using the model of Cropper et al
(1999) and in the right hand panel the fit using the same model plus 3
Gaussian components near 21.8 \AA which were used to model the O VII
line complex.}
\label{rgs_spec}
\end{figure*}

We extracted first order source and background spectra from each
detector using {\tt rgsproc}. We co-added the spectra from both
detectors using the procedure outlined by Page et al (2003) and then
binned the spectrum so that it had a minimum of 40 counts per bin. We
fitted the spectrum using the stratified accretion column model of
Cropper et al (1999). Since the magnetic field strength of the white
dwarf in V2301 Oph is low, we fixed the ratio of cooling due to
cyclotron and bremsstrahlung emission at the shock front
($\epsilon_{s}$) at a low value of 0.1. Further, we fixed the specific
mass accretion rate, $\dot{m}$, at 1.0 g s$^{-1}$ cm$^{-2}$, the metal
abundance at solar and $M_{1}$=1.0 \Msun. The spectra and best fit
(\rchi=1.01, 158 dof) to this model are shown in Figure \ref{rgs_spec}
(left).

The most prominent emission lines are those of the O VIII complex
between 18.9--19.3\AA\hspace{1mm} (0.642--0.655 keV) and the O VII
triplet between 21.6--22.1\AA \hspace{1mm} (0.561--0.574 keV). Less
prominent emission is also seen from Mg XI (8.42 \AA), Mg XII
(9.2--9.3 \AA) and Ne IX (or Fe XXII/XXIII). The accretion model of
Cropper et al (1999) fits the O VIII emission at $\sim$19.0\AA
\hspace{1mm} well, but not the O VII lines near 21.8 \AA. We then
added 3 Gaussians to our model to fit the O VII triplet and fixed them
at their expected rest energies (21.602, 21.804, 22.100 \AA)
\hspace{1mm} and set their width to be zero. We found that we obtained
a fit (\rchi=0.97, 155 dof) which was better than the previous fit at
a confidence of 97.4\%. In principal we can use the fitted parameters
of the Gaussian lines to the O VII complex to determine plasma
diagnostic ratios. However, the errors to the fit were too large to be
conclusive.

Ramsay et al (2004a) noted the presence of residuals in spectra
extracted from EPIC spectra of polars at this energy and suggested
that they may be due to the presence of an O VII emission line. Since
V2301 Oph is one of the brightest polars in the {\xmm}-MSSL polar
survey this is the first time that we have been able to confirm the
presence of this line in a spectrum taken using the RGS of a polar in
our sample.

The accretion column model of Cropper et al (1999) sums up the
contribution of optically-thin thermal plasma models (MEKAL) according
to the prescription of Aizu (1973) and includes the effects of
cyclotron cooling and the change in gravitational acceleration over
the height of the shock. It therefore does not include emission due to
photo-ionisation. We note that thermal plasma with very low
temperatures ($<0.5$keV) does emit O VII lines near 21.8 \AA, but when
the emission from hotter parts of the shock are included the line
emission from O VII near 21.8 \AA \hspace{1mm} is negligible.

Comparing the RGS spectrum of V2301 Oph with other cataclysmic
variables presented in Mukai et al (2003) the relative strength of
emission lines due to O VII compared with the lines due to O VIII is
similar to those systems which these authors identified as having
spectra dominated by photo-ionisation. The fact that the accretion
column model of Cropper et al (1999) fits the RGS spectrum well, with
the exception of wavelengths centered on the O VII line, suggests that
for polars their X-ray spectrum maybe best modelled with a
contribution from photo-ionised line emission to the collisionally
ionised plasma. This is to be expected since some fraction of the
X-ray photons emitted in the post-shock flow will intercept the
accretion stream resulting in emission of X-rays produced via
photo-ionisation.

\subsection{EPIC spectra}
\label{epic}

The light curves shown in the previous section suggest that there are
two distinct accretion regions, one of which appears significantly
softer than the other. Since the signal-to-noise of the EPIC data were
much higher than the RGS data, we were able to extract phase resolved
spectra from the EPIC data. We therefore initially extracted two
spectra, one taken from the bright region ($\phi$=0.05--0.2)
immediately after the deep eclipse (so that absorption effects were
minimised) and one from the fainter pole when the softness ratio was
significantly higher ($\phi$=0.3--0.4). We note the `bright' accretion
region spectrum will include photons from the `faint' accretion
region.

Most polars have X-ray spectra which have a strong soft X-ray
component which is produced as a result of re-processing of hard X-rays
from the photosphere of the white dwarf and is typically modelled
using a blackbody component with $kT_{bb}\sim$40eV (eg Ramsay et al
1994, Beuermann \& Burwitz 1995 and references therein). Our initial
fits to both the bright and faint regions showed that a soft blackbody
component was not required to give good fits to the X-ray
spectra. V2301 Oph therefore joins 7 other polars which have at least
one accretion pole which does not show a soft X-ray component (Ramsay
\& Cropper 2004). Ramsay \& Cropper (2004) proposed that these systems
do have a soft X-ray component, but have a temperature low enough so
that the blackbody is shifted to lower energies and therefore not seen
using X-ray detectors which typically have low energy cut-offs near
0.1keV. We discuss this further in the next section.

In our analysis of the EPIC spectra we again used the accretion model
of Cropper et al (1999) and fixed $\epsilon_{s}$=0.1 (as done for our
analysis of the RGS spectra). We initially set $\dot{m}$=1.0 g
s$^{-1}$ cm$^{-2}$. We included a neutral absorption model and a
Gaussian emission line fixed at 0.571 keV (= 21.7\AA) with zero width
to include the O VII line due to photo-ionisation (\S \ref{rgs}).

We begin our discussion with the spectra extracted from the bright
accretion region. We show the fits and spectral parameters in Table
\ref{epic_fits}. Since the EPIC pn data is slightly piled up at the
highest count rates, it is not surprisingly that the fits to the EPIC
MOS data are marginally better than the EPIC pn data. Fixing the metal
abundance at solar we find $M_{1}=1.1-1.2$\Msun. Freeing the metal
abundance gives masses which are lower by up to 0.1 \Msun.

From our analysis of the eclipse light curves we determined an upper
limit on $M_{1}$ of 1.2 \Msun (\S \ref{eclipse}). Our derived masses
assuming a specific mass accretion rate of 1.0 g s$^{-1}$ cm$^{-2}$
and solar abundance are therefore close to this upper limit. For a
metal abundance less than solar the masses are more consistent with
the results of the eclipse results. We have also examined the case of
$\dot{m}=0.1$ g s$^{-1}$ cm$^{-2}$, performing the same analysis as
before. For a range of metal abundance we find $M_{1}$=1.0--1.1\Msun.

For the faint spectra we fixed $M_{1}$=1.05 \Msun and $\dot{m}$ at 1.0
and 0.1 g s$^{-1}$ cm$^{-2}$.  We obtain good fits for both values of
$\dot{m}$ and show the fits in Table \ref{epic_fits}. This suggests
that the derived parameters for the faint pole spectra are not
sensitive to the specific mass transfer rate within the errors of the
fit.

We show the observed flux, the unabsorbed flux and the luminosity
(assuming a distance of 150 pc, Silber et al 1994) using the fits to
the EPIC MOS data in Table \ref{flux}. Assuming that the X-rays are
optically thin we find the luminosity for the bright pole is
2$\times10^{32}$ \ergss and for the faint pole 1.8$\times10^{31}$
\ergss. This compares with a mean value of 2.0$\times10^{32}$ \ergss
for the sample of polars in a high state observed using {\xmm} (Ramsay
\& Cropper 2004). This suggests that the bright pole has an X-ray
luminosity typical of polars. The fact that V2301 Oph shows the
highest count rate of any of the polars in our sample makes it likely
that it is one of the closest objects in our survey. To test this we
searched the literature to obtain distances to all the polars included
in the Ramsay \& Cropper (2004) sample (we note that many are without
even approximate limits to their distance). The only polar with a
lower limit to its distance closer than 150 pc is GG Leo (Ramsay et al
2004) which has a lower limit of 100 pc (Burwitz et al 1998), and
shows a peak count rate of $\sim$3 ct/s in the 0.15--10keV energy band
in the EPIC pn.

\begin{table*}
\begin{center}
\begin{tabular}{lrrrrr}
\hline
Detector & \multicolumn{5}{c}{Bright Pole} \\
         & $N_{H}$ & $M_{1}$ & Z & $\dot{m}$ & \rchi \\
         & ($10^{20}$ \pcmsq) & (\Msun) & (Solar) & g cm$^{-2}$ s$^{-1}$ &
(dof) \\
\hline
EPIC pn  & 4.2$\pm$0.4 & 1.17$\pm$0.04 & 1.0 & 1.0 & 1.17 (286)\\
EPIC MOS & 4.8$\pm$0.7 & 1.19$\pm$0.03 & 1.0 & 1.0 & 1.05 (198)\\
EPIC pn  & 4.5$^{+0.7}_{-0.5}$ & 1.14$^{+0.03}_{-0.08}$ & 
0.50$^{+0.22}_{-0.25}$ & 1.0 & 1.12 (285)\\
EPIC MOS & 5.7$\pm$0.6 & 1.10$\pm$0.09 & 0.15$^{+0.2}_{-0.15}$ &
1.0 & 0.97 (197) \\
EPIC pn  & 4.3$\pm$0.4 & 1.06$\pm$0.04 & 1.0 & 0.1 & 1.17 (286)\\
EPIC MOS  & 4.9$^{+0.8}_{-0.6}$  & 1.09$\pm$0.05 & 1.0 & 0.1 & 1.03 (198)\\
EPIC pn  & 4.3$\pm$0.4 & 1.06$\pm$0.04 & 0.5$\pm$0.2 & 0.1 & 1.17 (285)\\
EPIC MOS  & 5.4$^{+0.9}_{-0.6}$ & 1.05$^{+0.04}_{-0.08}$ & 
0.3$^{+0.2}_{-0.3}$ & 0.1 & 0.96 (197)\\
\hline
Detector & \multicolumn{5}{c}{Faint Pole} \\
         & $N_{H}$ & $M_{1}$ & Z & $\dot{m}$ & \rchi \\
         & ($10^{20}$ \pcmsq) & (\Msun) & (Solar) & g cm$^{-2}$ s$^{-1}$ &
(dof) \\
\hline
EPIC pn  & 2.6$^{+1.8}_{-1.5}$ & 1.05 & 1.0 & 1.0 & 0.98 (93)\\
EPIC MOS  & 3.2$^{+1.6}_{-1.2}$ & 1.05 & 1.0 & 1.0 & 0.88 (125)\\
EPIC pn  & 2.2$^{+1.8}_{-1.3}$ & 1.05 & 1.0 & 0.1 & 1.05 (93)\\
EPIC MOS  & 2.6$^{+1.6}_{-1.3}$ & 1.05 & 1.0 & 0.1 & 0.91 (125)\\
\hline
\end{tabular}
\end{center}
\caption{The parameters for the spectral fits to the bright and faint
poles seen in V2301 Oph. We used a stratified accretion column model
which includes the changing gravitational potential over the height of
the accretion shock (Cropper et al 1999). We show the fit from data
extracted from the EPIC pn detector and also the combined fit to data
extracted from the EPIC MOS 1 and 2 detectors.  In one fit we fixed
the metal abundance at solar and in the another we allowed it to
vary. We fixed the specific accretion rate at 1.0 and 0.1 g s$^{-1}$
cm$^{-2}$. Errors are given at the 90\% confidence level.}
\label{epic_fits}
\end{table*}

\begin{table}
\begin{center}
\begin{tabular}{l@{\hspace{5pt}}r@{\hspace{5pt}}r@{\hspace{5pt}}r}
\hline
    &  Observed flux & Unabsorbed Flux & $L_{X}$ \\
    & (0.2-10keV)    & (Bolometric) &  (Bolometric) \\ 
    & $\times10^{-12}$ & $\times10^{-12}$ & $\times10^{31}$ \\
    & \ergscm        & \ergscm & \ergss \\
\hline
Bright Pole & 31$^{+1}_{-2}$ & 74$^{+3}_{-3}$ & 20$\pm1$ \\ 
Faint Pole & 3.2$^{+0.3}_{-0.3}$ & 6.6$^{+0.6}_{-0.6}$ & 1.8$\pm0.1$ \\ 
\hline
\end{tabular}
\end{center}
\caption{Using the fits to the EPIC MOS spectra, we show the observed 
flux in the 0.2-10keV energy band, the unabsorbed bolometric flux and the 
unabsorbed bolometric luminosity for the bright and faint poles assuming 
a distance of 150 pc (Silber et al 1994).} 
\label{flux}
\end{table}

\section{The accretion region in V2301 Oph}

We found from our X-ray spectral fits that our accretion models
predict $M_{1}$=1.1--1.2 \Msun when we assumed $\dot{m}$ = 1.0 \gscm
and $M_{1}$=1.0--1.1 \Msun when we assumed $\dot{m}$ = 0.1 \gscm.  Our
results from the eclipse profile showed that $M_{1}$=1.2 \Msun was an
upper limit. The specific accretion rate cannot therefore not be much
greater than 1.0 \gscm.

What does this imply for the fractional area that accretion is
occurring over? In \S \ref{epic} we found $L_{X}=2\times10^{32}$
\ergss. If all the accretion energy is emitted as X-rays, this implies
a mass transfer rate of $8\times10^{14}$ g/s. For $M_{1}$=1.0\Msun
this implies that accretion is occurring over a fraction, $f$, of 0.002
and 0.0002 of the white dwarf surface for $\dot{m}$=0.1 and 1.0 \gscm
respectively. (If a significant amount of accretion energy is emitted
at other wavelengths then this fraction is a lower limit). The
fraction implied for $\dot{m}$=0.1 \gscm is larger than normally seen
in polars but similar to that found in intermediate polars (eg Rosen
1992, James et al 2002) - these binaries have an accreting white dwarf
whose magnetic field strength is not high enough to synchronise the
spin of the white dwarf with the orbital period.

We noted earlier that many polars show a distinct soft X-ray component
which is caused by re-processing of hard X-ray emitted in the
post-shock flow by the photosphere of the white dwarf. V2301 Oph shows
no evidence for this soft X-ray component and therefore joins 6 other
polars with this feature. The temperature of the reprocessed component
is proportional to $(\dot{M}/f)^{1/4}$ where $\dot{M}$ is the total
mass transfer rate (eg King \& Lasota 1990). Therefore for large
values of $f$ and low values of $\dot{M}$ the effective temperature,
$T_{eff}$ will be lower. If this is sufficiently low this component
will move out of the soft X-ray band and into the far UV.

Most intermediate polars do not show a distinct soft X-ray
component. V2301 Oph which has the lowest magnetic field strength of
any polar is consistent with this characteristic. Of the 7 systems
which do not show a soft component 3 (BY Cam, RX J2115--58 and V1500
Cyg) are asynchronous polars (those polars in which the spin period of
the accreting white dwarf and the binary orbital period differ by a
few percent) which may imply that accretion flow is spread over a
longer range of magnetic azimuth than normal. The specific accretion
rate might therefore be too low over a significant percentage of the
accretion region to produce a strong shock which will affect the
presence and temperature of any reprocessed component.

\section{Absorption Dip}

We extracted EPIC spectra taken from time intervals corresponding to
the orbital phase during which there was the pre-eclipse absorption
dip, and also time intervals just preceding this dip (which we call
the `dip-free' spectrum). The dip feature corresponds to the point in
the binary orbital cycle where the accretion stream passes directly
between the observer and the bright accretion region(s) on the
accreting white dwarf for a wide range of viewing angles and magnetic
field orientations.

We modelled the dip-free spectrum using the model described in \S
\ref{epic}. We find that the absorption column for the neutral
absorber is 4$\times10^{20}$ \pcmsq (consistent with that shown in
Table \ref{epic_fits}). In modelling the pre-eclipse absorption dip
spectrum we therefore fixed the absorption of this component at this
value. As expected, the fits to the EPIC spectrum using this model was
poor. We therefore added a second absorption model to account for
stream absorption. For the latter we chose the {\tt absori} and {\tt
pcf} models in {\tt XSPEC}. The former is a warm absorber which
assumes an emission model which can be represented by a power law
(here fixed at a slope of 1.4 which correctly fits the bright phase
X-ray spectrum) and the latter is a partial covering model. Using the
former model we obtain a total absorption column for the accretion
stream $N_{H}=2.9\times10^{21}$ \pcmsq (\rchi=0.83, 447 dof), and the
latter model $N_{H}=3.2\times10^{21}$ \pcmsq, with a partial covering
of 0.95, (\rchi=0.84, 447 dof). This absorption column density is
similar to that found for RX J1002--19 but lower than in EV UMa
($7-20\times10^{22}$ \pcmsq, Ramsay \& Cropper 2003).

\section{Conclusions}

V2301 Oph shows a unique standstill in its X-ray light curve.  This
allows us to place an upper limit of $M_{1}=$1.2 \Msun if we assume
that the X-rays originate close to the white dwarf surface. Being the
polar with the lowest magnetic field strength (7MG), V2301 Oph is the
best target to detect QPO's in the X-ray flux since higher field
strengths are predicted to suppress oscillations.  Our non-detection
of QPO's in X-rays suggest that either QPO's are suppressed even for
magnetic field strengths of 7MG or that they would only be detected at
higher energies which are expected to be emitted closer to the
accretion shock front.  We find that V2301 Oph joins 6 other polars
which do not show evidence for a distinct soft X-ray component. We
believe that this is due to the fact that the temperature of the
reprocessed component is low enough to be shifted out of the {\xmm}
pass band. We speculate that this could be due to the relatively high
fractional area (derived from our fits to the EPIC spectra) that
accretion is occurring on the white dwarf.

\section{acknowledgments}

This is work based on observations obtained with {\sl XMM-Newton}, an
ESA science mission with instruments and contributions directly funded
by ESA Member States and the USA (NASA). These data were obtained as
part of the OM consortium guaranteed time.  We thank Ali Kinkhabwala
and Curtis Saxton for useful discussions.

\end{document}